\documentclass[a4paper,fleqn,titlepage,twocolumn,superscriptaddress,showkeys]{revtex4-2}
\usepackage[utf8]{inputenc}
\usepackage{graphicx}
\usepackage{amsmath}
\usepackage{amssymb}
\usepackage{times}
\usepackage[utf8]{inputenc}
\usepackage{nicefrac}
\usepackage{color}
\usepackage{xcolor}
\usepackage{hyperref}
\hypersetup{
     colorlinks=true,
     linkcolor=blue,
     filecolor=blue,
     citecolor = blue,      
     urlcolor=blue}
     
\usepackage{color}
\usepackage{subcaption}
\usepackage{dsfont}
\usepackage[ruled,vlined]{algorithm2e}
\usepackage{xr} 
\graphicspath{{figs/}}

\newcommand{\eg}{\emph{e.g.}\ }
\newcommand{\ie}{\emph{i.e.}\ }
\newcommand{\cf}{\emph{c.f.}\ }
\newcommand{\resp}{\emph{resp.}\ }

\newtheorem{lemma}{Lemma}

\newcommand{\gald}{\langle\rho\rangle}

\begin{document}


\title{Discord in the voter model for complex networks}

\author{Antoine Vendeville}
\email{Corresponding author: a.vendeville@ucl.ac.uk}
\affiliation{Department of Computer Science, University College London, UK}
\author{Shi Zhou}
\affiliation{Department of Computer Science, University College London, UK}
\author{Benjamin Guedj}
\affiliation{Department of Computer Science, University College London, UK} 
\affiliation{Inria Lille - Nord Europe, 59650 Villeneuve d'Ascq, France}

\date{\today}

\begin{abstract}
	Online social networks have become primary means of communication. As they often exhibit undesirable effects such as hostility, polarisation or echo chambers, it is crucial to develop analytical tools that help us better understand them. In this paper, we are interested in the evolution of discord in social networks. Formally, we introduce a method to calculate the probability of discord between any two agents in the multi-state voter model with and without zealots. Our work applies to any directed, weighted graph with any finite number of possible opinions, allows for various update rates across agents, and does not imply any approximation. Under certain topological conditions, their opinions are independent and the joint distribution can be decoupled. Otherwise, the evolution of discord probabilities is described by a linear system of ordinary differential equations. We prove the existence of a unique equilibrium solution, which can be computed via an iterative algorithm. The classical definition of active links density is generalized to take into account long-range, weighted interactions. We illustrate our findings on real-life and synthetic networks. In particular, we investigate the impact of clustering on discord, and uncover a rich landscape of varied behaviors in polarised networks. This sheds lights on the evolution of discord between, and within, antagonistic communities.
\end{abstract}

\maketitle

\section{Introduction} 
There is growing concern over the nefarious effects of online social networks. One phenomenon in particular, has attracted more and more attention. Similar-minded users tend to gather, and form strongly opinionated communities: echo chambers. This is particularly salient in polarised debates regarding politics, societal issues or conspiracy theories \cite{williams2015,bakshy2015,garimella2018,cinelli2021,kirdemir2022}. Echo chambers tend to foster animosity between opposite sides and fuel reinforcement of pre-existing beliefs \cite{dandekar2013,spohr2017}---hence their name. To better understand the phenomenon, it is crucial to first examine how \emph{discord}---\ie disagreement---evolves in social networks. In this paper we contribute novel insight towards this goal. In the context of the celebrated Voter Model, we demonstrate a formula to calculate discord probabilities on complex networks, defined as
\begin{equation}
	\rho_{ij} := P(\sigma_i\neq\sigma_j),
\end{equation}
where $\sigma_i$ is the opinion of agent $i$ (\resp\ $\sigma_j$, agent $j$).

In the Voter Model, agents within a graph are endowed with individual binary opinions $\{0,1\}$. They proceed to choose a random neighbor and adopt their opinion, repeating this process multiple times. On the $\mathbb{Z}^d$ lattice $(d\le 2)$ as well as on any finite connected network, this dynamic leads to a state of consensus where all agents agree \cite{frachebourg1996,yildiz2010}. The average discord $\gald$ over all edges, commonly referred to as \emph{active links density}, is an order parameter of great interest. A low active links density indicates the presence of large clusters wherein all agents agree, and disagreement is only found at their borders. On the other hand, when it is high, there are no such structures and opinions are spread uniformly on the network.

In finite uncorrelated networks, $\gald$ typically decays exponentially fast, and the convergence time scales linearly with $N$ \cite{castellano2005,vazquez2008,pugliese2009}. Similar results have been obtained for small-world networks, while scale-free networks present a different behaviours depending on the exponent of the power-law degree distribution \cite{pugliese2009,carro2016,peralta2018,ramirez2022}. However, most of these rely on approximations, and no general solution applicable to any given network has been found. If we wish to study the active links density in real-world networks, this is of primary importance.

In this paper, we present a universal formula for discord probabilities that applies to any given directed, weighted network, without any requirement regarding the topology or the degree distribution of the network. Our framework can accomodate any finite number of opinions \cite{yildiz2010,starnini2012}, zealots \cite{chinellato2015,mobilia2007}, and individual update rates \cite{masuda2010}. Importantly, we make no approximation and the results are exact. We also propose a new way of computing the active links density, that uses the exponential of the adjacency matrix to account for long-range, weighted interactions.

The rest of this manuscript is organised as follows. Section~\ref{system} introduces the model. In Sect.~\ref{equilibrium_opinions} we derive equations for the evolution of individual opinion distributions. In Sect.~\ref{discord_section} we present our main results: the formula for discord probabilities and the generalized active links density. Section~\ref{experiments} is devoted to numerical experiments that demonstrate the use cases of our results. Finally, we conclude in Sect.~\ref{conclusion}.

\section{System} \label{system}
Consider a group $\mathcal{N}=\{1,\ldots,N\}$ of agents interacting through a social network and holding discrete opinions $\sigma_1, \ldots, \sigma_N$ valued in some set $\mathcal{S}=\{1,\ldots,S\}$. Opinions are bound to evolve due to influence from peers, and we let $w_{ij}\ge 0$ denote the weight of the edge $j\rightarrow i$. Agents may also be under the influence of zealots, one-sided sources that never change opinions. Possible examples include strongly opinionated characters, such as politicians, journalists or lobbyists.  

As in \cite{masuda2015,khalil2018}, we consider all zealots with the same opinion $s$ as one single entity exogenous to the network, which we name the $s$-zealot. The word \emph{agent} will specifically refer to non-zealous individuals. In this case, the notion of zealots has a broader interpretation as exogenous sources of influence, and may for example describe some inner bias of the agents, or an advertising campaign. We let $z_i^s$ denote the total amount of influence exerted on agent $i$ by the $s$-zealot. We say that agent $i$ can be influenced by the $s$-zealot if either $z_i^s>0$ or there exists an agent $j$ with $z_j^s>0$ and a path from $j$ to $i$. For the latter, the strength of influence exerted by the $s$-zealot on $i$ will depend on the cumulated length of all such paths. We discuss long-range influences in more details in Sect.~\ref{gald}.

We call \emph{leaders of} $i$ agents in the set $\mathcal{L}_i=\{j\in\mathcal{N}:w_{ij}>0\}$. We allow self-loops, \ie $i\in\mathcal{L}_i$. They can be used to describe \emph{(i)} an unwillingness of agents to reevaluate their opinions, or \emph{(ii)} agents re-affirming their current opinions. The former can effectively adjust the update rates. The latter can be useful when using the Voter Model to describe the dynamics of people retweeting their own previous posts. 

For the sake of simplicity and without loss of generality, we consider edge weights to be normalised: for any agent $i\in\mathcal{N}$,
\begin{equation} \label{normalisation}
	\sum_{j\in\mathcal{L}_i} w_{ij}+\sum_{s\in\mathcal{S}}z_i^s=1.
\end{equation}
In particular, each node is assumed to either have a leader or to be connected to a zealot. The directed, weighted graph of all agents is denoted by $\mathcal{G}$. We assume $\mathcal{G}$ to be weakly connected---if it is not, one can apply the results to each component of the graph separately. We let $W=(w_{ij})_{1\le i,j\le N}$ denote the weighted adjacency matrix of the graph. 

\section{Evolution of opinions} \label{equilibrium_opinions}
Each agent is endowed with an exponential clock of parameter 1. Whenever $i$'s clock rings, $i$ updates their opinion by copying a random neighbor or a zealot. Neighbor $j$ is copied with probability $w_{ij}$, and the $s$-zealot with probability $z_i^{(s)}$. In other words, with rate $w_{ij}$ agent $i$ copies agent $j$, and with rate $z_i^s$ agent $i$ adopts opinion $s$ via the $s$-zealot. 

Let $x_i^s$ denote the probability for agent $i$ to hold opinion $s$. To avoid cumbersome notations, we omit the time parameter. We shall remember that $x_i^s$ is a time-dependent quantity and is implicitly followed by $(t)$. A direct extension of \cite[eq.\ 3]{masuda2015} gives
\begin{align}
\frac{dx_i^s}{dt} = &(1-x_i^s) \left[ \sum_{j\in\mathcal{L}_i} w_{ij}x_j^s + z_i^s \right] \\ &-x_i^s \left[ \sum_{j\in\mathcal{L}_i} w_{ij} (1-x_j^s) + \sum_{s^\prime\neq s} z_i^{s^\prime} \right],
\end{align}
which reduces to
\begin{align}
	\frac{dx_i^s}{dt} &= \sum_{j\in\mathcal{L}_i} w_{ij}x_j^s +z_i^s -x_i^s. \label{dxdt}
\intertext{Hence, at equilibrium we have}
	x_i^s &= \sum_{j\in\mathcal{L}_i} w_{ij}x_j^s +z_i^s. \label{xstar_eq}
\end{align}
As expressed by \cite[Prop.~3.2]{yildiz2013}, $x_i^s$ is the probability that a backward random walk initiated at $i$ reaches the $s$-zealot before another zealot. Hence, $x_i^s>0$ if and only if $i$ can be influenced by the $s$-zealot. 

The dynamics correspond to those of a continous-time Friedkin-Johnsen model \cite{friedkin_johnsen}. The number of distinct equilibrium states depends on the topology of the agent graph and the influence of zealots. Assuming that any agent can be influenced by at least one zealot, Eq.~\ref{normalisation} and \cite[Lemma 2.1]{azimzadeh2018} imply that the spectral radius of $W$ is strictly less than 1, hence there is a unique equilibrium state---\cf\ Appendix~\ref{unicity}. If $z_i^s=0$ for all $i,s$, we uncover the continuous-time French-DeGroot model \cite{french,degroot}. In that case, consensus is reached if there exists an agent able to reach every other. 

\section{Discord probabilities} \label{discord_section}
\begin{figure}
     \centering
     \begin{subfigure}[b]{.2\textwidth}
         \centering
         \includegraphics[width=\textwidth]{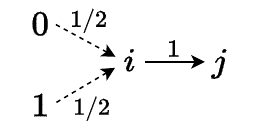}
		 \caption{Path $i\rightarrow j$.}
     \end{subfigure}~
     \begin{subfigure}[b]{.2\textwidth}
         \centering
         \includegraphics[width=\textwidth]{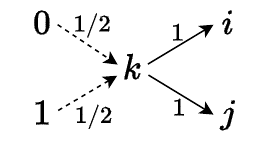}
		 \caption{Common ancestor $k$.}
     \end{subfigure}
     \caption{Dependency between opinions. Nodes $0$ and $1$ are the 0- and 1-zealot respectively. Numbers along the arrows denote edge weights. In (a) there is a path from $i$ to $j$. In (b) there is none, but $i$ and $j$ have a common ancestor $k$. In both cases, Eq.~\ref{qij_dep} gives $\rho_{ij}=1/4$, while Eq.~\ref{qij_indep} gives $\rho_{ij}=1/2$.}
     \label{counter_ex}
\end{figure}

Let $\rho_{ij}$ denote the probability $P(\sigma_i\neq \sigma_j)$ of discord between agents $i$ and $j$. The quantity $1-\rho_{ij}$ is called probability of \emph{harmony}. Trivially $\rho_{ii}=0$ and $\rho_{ij}=\rho_{ji}$. For the sake of conciseness we again omit the time parameters, and denote indistinctively by $ij$ or $ji$ the unordered agent pair $\{i,j\}$. One can then write
\begin{align}
	\rho_{ij} &= P(\sigma_i\neq \sigma_j) \\
	&= \sum_{s\in\mathcal{S}} P(\sigma_i=s,\sigma_j\neq s) \\
	&= \sum_{s\in\mathcal{S}} P(\sigma_i=s)P(\sigma_j\neq s). \label{indep_eq} \\
	&= \sum_{s\in\mathcal{S}} x_i^s(1-x_j^s).\label{qij_indep}
\end{align} 
However, the above derivation implictly assumes that the opinions $\sigma_i$ and $\sigma_j$ are independent, which is not guaranteed---\eg if $i$ and $j$ are neighbors. A toy example is presented in Fig.~\ref{counter_ex}. We focus on the general case, valid for any $i,j$. Later on we characterise precisely the cases where Eq.~\ref{qij_indep} holds.

\subsection{General case}
There are two types of events that lead to $i$ adopting an opinion different than $j$'s:
\begin{enumerate}
	\item $i$ copies agent $k\neq j$, who holds another opinion than $j$'s. This happens at rate $w_{ik}\rho_{jk}$.
	\item $i$ copies an $s$-zealot while $j$ holds an opinion different than $s$. This happens at rate $z_i^s(1-x_j^s)$.
\end{enumerate}
Hence, $i$ adopts another opinion than $j$'s at rate 
\begin{equation}
	\sum_{k\in\mathcal{L}_i} w_{ik}\rho_{jk} + \sum_{s\in\mathcal{S}} z_i^s(1-x_j^s).
\end{equation}
The same reasoning gives us the rate at which $j$ adopts another opinion than $i$'s, and we find that the pair $ij$ switches from harmony to discord at rate:

\begin{widetext}
\begin{align}
	\Delta_{ij}^- = (1-\rho_{ij}) &\left[ \sum_{k\in\mathcal{L}_i} w_{ik}\rho_{jk} + \sum_{s\in\mathcal{S}} z_i^s(1-x_j^s) +\sum_{k\in\mathcal{L}_j} w_{jk}\rho_{ik} + \sum_{s\in\mathcal{S}} z_j^s(1-x_i^s) \right],
\intertext{and from discord to harmony at rate}
	\Delta_{ij}^+ = \rho_{ij} &\left[ \sum_{k\in\mathcal{L}_i} w_{ik}(1-\rho_{jk}) + \sum_{s\in\mathcal{S}} z_i^sx_j^s +\sum_{k\in\mathcal{L}_j} w_{jk}(1-\rho_{ik}) + \sum_{s\in\mathcal{S}} z_j^sx_i^s \right].
\intertext{Subtracting $\Delta_{ij}^-$ from $\Delta_{ij}^+$, we obtain the master equation}
 	 \frac{d\rho_{ij}}{dt} = \Delta_{ij}^+ - \Delta_{ij}^- =&\sum_{k\in\mathcal{L}_i} w_{ik}\rho_{jk} + \sum_{k\in\mathcal{L}_j} w_{jk}\rho_{ik} +\sum_{s\in\mathcal{S}} z_i^s(1-x_j^s) + \sum_{s\in\mathcal{S}} z_j^s(1-x_i^s) -2\rho_{ij}. \label{master_eq}
\intertext{The evolution of discord probabilities is thus governed by a system of $N(N-1)/2$ linear differential equations---although not all will be needed, as stressed below. Setting the left-hand side to zero gives us the equilibrium discord probability for the pair $ij$,}
\rho_{ij} = \frac{1}{2} &\left[ \sum_{k\in\mathcal{L}_i} w_{ik}\rho_{jk} + \sum_{k\in\mathcal{L}_j} w_{jk}\rho_{ik} + \sum_{s\in\mathcal{S}} z_i^s(1-x_j^s) + \sum_{s\in\mathcal{S}} z_j^s(1-x_i^s)\right].  \label{qij_dep}
\end{align}
\end{widetext}
The discord probabilities are thus solution of a linear system of the form 
\begin{align}
	\rho &= V\rho+y, \quad &\text{(equilibrium)} \label{linearsystem} \\
	\frac{d\rho}{dt} &= 2(V-I)\rho+2y.  \quad &\text{(evolution)}
\end{align}
The entries of $V,y$ are given by \eqref{qij_dep}. In the absence of zealots, we have $y=0$ and the speed of convergence depends on the spectrum of $V$. If consensus is reached (\eg strongly connected network), then $\rho\rightarrow 0$. Otherwise, the various equilibrium states are given by the leading eigenvectors of $V$.

In the presence of zealots, Lemma 2.1 from \cite{azimzadeh2018} implies that the spectral radius of $W$ is strictly less than 1, assuming every agent can be influenced by a zealot---\cf\ Appendix~\ref{unicity} for details. Hence, the system has a unique solution. It can be efficiently computed by iterating 
\begin{equation}\label{iterate}
	\rho^{(k)}=V\rho^{(k-1)}+y
\end{equation}
for any initialisation $\rho^{(0)}$ with values in $]0,1[$, and the convergence rate depends on the spectral radius of $V$. The proof of that statement can be found in \cite[Thm.~4]{giovanidis2021}. 

Finally, note that all the equations we derived can be adapted to account for various update rates $r_1, \ldots, r_N$ across agents. It suffices to scale the rates $w_{ij}$ by $r_i$, to replace $2\rho_{ij}$ by $(r_i+r_j)\rho_{ij}$ in Eq.~\ref{master_eq} and $1/2$ by $1/(r_i+r_j)$ in Eq.~\ref{qij_dep}. Equations~\ref{dxdt}-\ref{xstar_eq} are unchanged. Here, we stick to the traditional setting $r_i=1$ for all $i$.

\subsection{Independent pairs} \label{indep_pairs}
There are some cases where the opinions $\sigma_i$ and $\sigma_j$ are independent, and one can use Eq.~\ref{qij_indep} to calculate $\rho_{ij}$ without having to solve a possibly large linear system. Independence holds if one of the following is verified:
\begin{enumerate}
	\item $\sigma_i$ or $\sigma_j$ is constant, or
	\item there is no path from $i$ to $j$ nor from $j$ to $i$, and $i$ and $j$ have no common ancestor.
\end{enumerate}
A formal proof can be found in the Appendix~\ref{independence}. The second condition assures us that $i$ and $j$ do not influence each other, and that they are exposed to strictly different sources: their opinions evolve in total independence. As illustrated in Fig.~\ref{counter_ex}, if one of these assumptions is violated  then Eq.~\ref{qij_indep} gives an incorrect result.

When $i$ and $j$ have the same opinion distribution $x$, Eq.~\ref{qij_indep} is akin to the entropy of $x$. Hence the more uniform $x$ is, the higher the discord. Equation~\ref{qij_indep} is also exactly 1 minus the cosine similarity between $x_i$ and $x_j$. While for dependent agent pairs the discord probabilities are given by Eqs.~\ref{master_eq} and \ref{qij_dep}, Eq.~\ref{qij_indep} may still be used for the purpose of measuring dissimilarity of opinion distributions.

\subsection{Generalized active links density} \label{gald}
The active links density is the average discord between all neighboring agents. While convenient for regular, unweighted graphs, this definition suffers from two shortcomings when it comes to general networks. First, not all edges are created equal: if $w_{ij}=0.80$ and $w_{ik}=0.01$, then $j$ holds a strong power of influence over $i$, while $k$ barely has any at all. Discord $\rho_{ij}$ between $i$ and $j$ will thus often be much more relevant to the analysis than $\rho_{ik}$, a difference not accounted for when taking a simple unweighted average. 

Second, two agents may be closer than they appear: if $w_{ij}=0$ but $w_{ik}=w_{kj}=0.9$, agent $j$ exerts non-negligible influence on $i$ via $k$, despite them not being directly connected by an edge. The opinion of agent $j$ reaches $i$ in two steps with rate
\begin{equation} \label{longer_route}
	\sum_{k\in\mathcal{L}_i} w_{ik}w_{kj}.
\end{equation}
Indeed $i$ copies leader $k$ with rate $w_{ik}$, and each time it happens there is a probability $w_{kj}$ that $k$'s opinion was copied directly from $j$.

From these considerations stems a novel metric, better suited for complex networks: the \emph{generalized active links density}, defined over all agent pairs by
\begin{equation} \label{ALD}
	\gald = \frac{\sum_{i<j} (w^\infty_{ij}+w^\infty_{ji})\rho_{ij}}{\sum_{i<j}(w^\infty_{ij}+w^\infty_{ji})}.
\end{equation}
Here, $w^\infty_{ij}$ can be any measure of influence of $j$ on $i$. Inspired by \cite{estrada2014}, here $w^\infty_{ij}$ is the $(i,j)$-th component of the matrix exponential
\begin{equation}
	e^W = \sum_{k=1}^\infty \frac{1}{k!} W^k.
\end{equation}
Summing all powers of $W$ generalises the reasoning made above \eqref{longer_route}, and lets us quantify the multi-steps influence that agents may have on one another. Scaling by the inverse of the path length factorial attributes a rapidly decaying importance to longer paths. The sum $w^\infty_{ij}+w^\infty_{ji}$ is then a measure of long-range, weighted influence between $i$ and $j$. 

Note that with this definition, zero entries in the $i^\text{th}$ row $w^\infty_{i}$ correspond to users unable to reach $i$, and non-zero entries correspond to ancestors of $i$. The cosine similarity
\begin{equation} \label{cosine}
	cos(w^\infty_{i},w^\infty_{j}) = \frac{w^\infty_{i}\cdot w^\infty_{j}}{\Vert w^\infty_{i}\Vert \Vert w^\infty_{j}\Vert}
\end{equation} 
informs us on the similarity of $i$ and $j$'s ancestry, meaning the extent to which they are exposed to the same channels of influence.

\section{Numerical experiments} \label{experiments}

\subsection{Dependency between opinions} \label{dependency}
One may be interested only in certain values of $\rho_{ij}$, and wishing to avoid the burden of computing them all. It seems natural to use 
\begin{equation}
	\tilde\rho_{ij} =  \sum_{s\in\mathcal{S}} x_i^s(1-x_j^s),
\end{equation}
as given by Eq.~\ref{qij_indep}, even if $\sigma_i$ and $\sigma_j$ are not independent. While sometimes effective, this approximation does not always fare well---\cf Fig.~\ref{counter_ex}. As the dependency of $i$ and $j$'s opinions relies on the strength of paths joining them and the similarity of their ancestry, we expect $\rho_{ij}$ to decrease with these, and the error made by $\tilde\rho_{ij}$ to increase. We verify this at equilibrium on four datasets: \texttt{zachary} \cite{zachary}, \texttt{football} \cite{football}, \texttt{email} \cite{email}, \texttt{polblogs} \cite{polblogs}. More details on the data and the preprocessing are available in the Appendix~\ref{datasets}. To quantify path strength and ancestry similarity we use respectively $w^\infty_{ij}+w^\infty_{ji}$ and $cos(w^\infty_{i},w^\infty_{j})$. We also look at the effect of the total zealousness $\Vert z_i+z_j\Vert$ of $i$ and $j$. 

\begin{figure}
     \centering
     \begin{subfigure}[t]{.1597\textwidth}
         \centering
         \includegraphics[width=\textwidth]{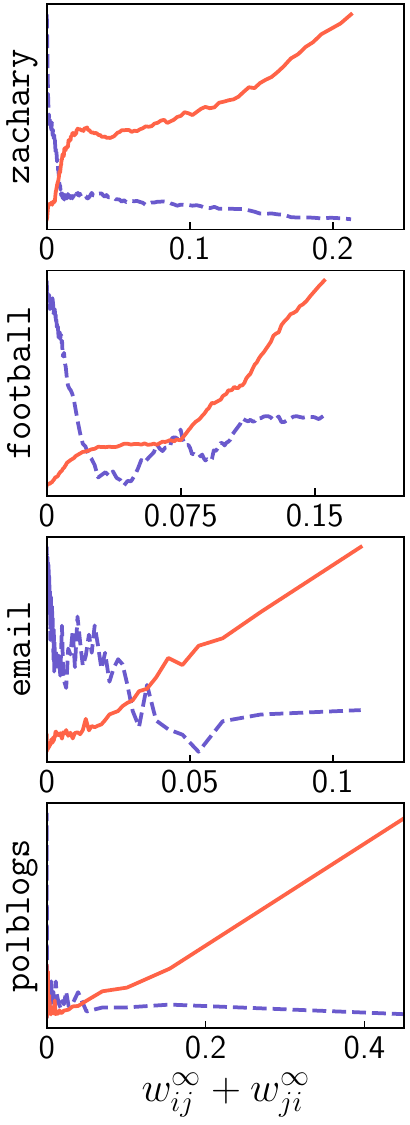}
		 \caption{Path strength}
     \end{subfigure}~
	 \begin{subfigure}[t]{.1445\textwidth}
		\centering
		\includegraphics[width=\textwidth]{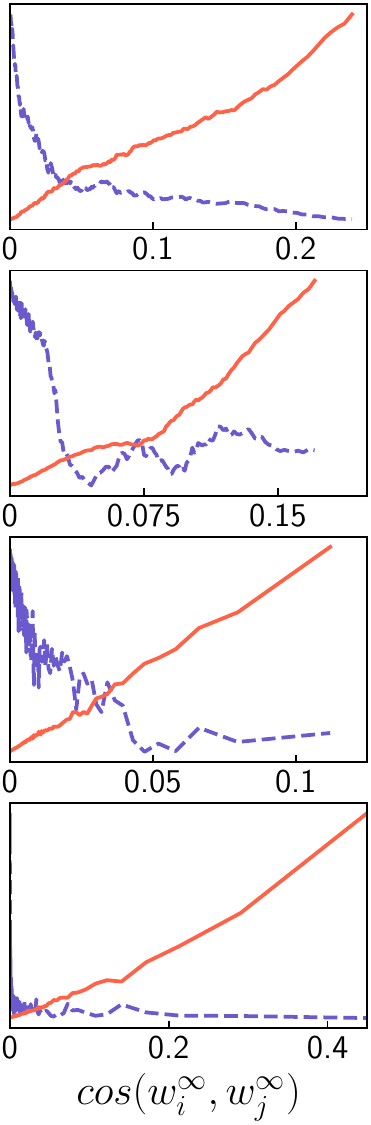}
		\caption{Ancestry similarity}
	\end{subfigure}~
	\begin{subfigure}[t]{.1597\textwidth}
		\centering
		\includegraphics[width=\textwidth]{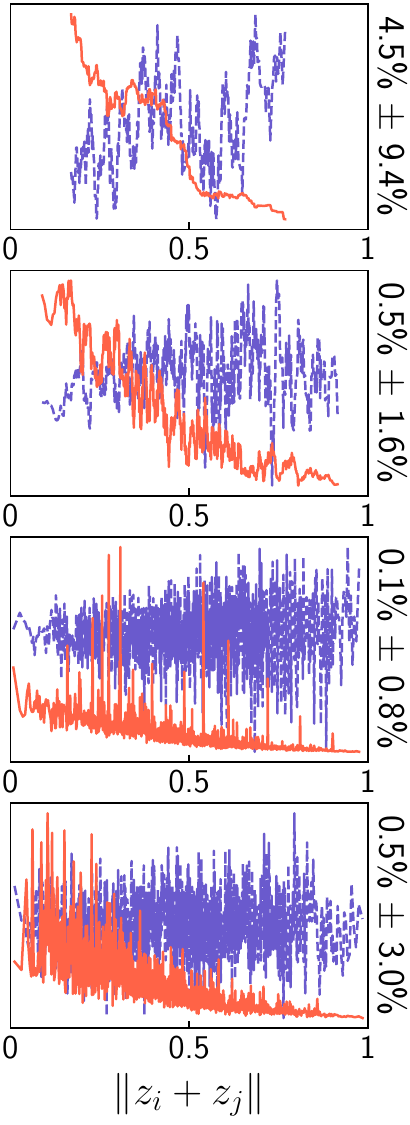}
		\caption{Total zealousness}
	\end{subfigure}\\
    \begin{subfigure}[b]{.4\textwidth}
        \centering
        \includegraphics[width=\textwidth]{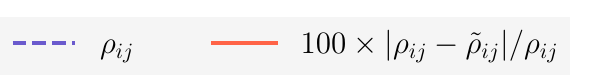}
    \end{subfigure}
    \caption{Moving average for $\rho_{ij}$ (dotted blue line) and error percentage of $\tilde\rho_{ij}$ (orange line) for all dependent agent pairs, function of \textbf{(a)} path strength, \textbf{(b)} ancestry similarity, and \textbf{(c)} and total zealousness. All curves differ in their range of values, which we don't precise on the y-axes for the sake of clarity. On the right are shown average errors and standard deviation for the distribution of errors.}
     \label{zpfe}
\end{figure}

Results are shown in Fig.~\ref{zpfe}, and confirm our hypotheses: $\rho_{ij}$  decreases with the strength of paths joining $(i,j)$ and the similarity of their ancestry, while the error made by $\tilde\rho_{ij}$ increases. Moreover, the error decreases with the total zealousness. This is not surprising, as higher values mean lighter weights on inter-agent edges and thus less influence from peers. The errors are quite low on average, but can peak very high for certain agent pairs (maximum error ranges from 15\% for \texttt{football} to 187\% for \texttt{polblogs}). The least accurate $\tilde\rho_{ij}$ is on \texttt{zachary}, which could be because of the higher path strengths due to the graph's smaller size.

\subsection{Discord and clustering} \label{smallworld} 
We investigate the relationship between discord and clustering. To do so, we compute the Generalized active links density $\langle\rho\rangle$ on a collection of Watts-Strogatz networks \cite{watts_strogatz}, with various average degrees and rewiring probabilities. There are $N=100$ agents arranged in a circle, and two possible opinions $\{0,1\}$. Each node is connected to its $k$ nearest neighbors, and each edge is rewired at random with probability $p$. As $p$ increases, the average local clustering coefficient $\langle C\rangle$---defined at node $i$ as the proportion of edges that exist among neighbours of $i$---decreases. Thus, there are less and less triangles in the graph.

\begin{figure*}
	\centering
	\begin{subfigure}[t]{.23\textwidth}
		\centering
		\includegraphics[width=\textwidth]{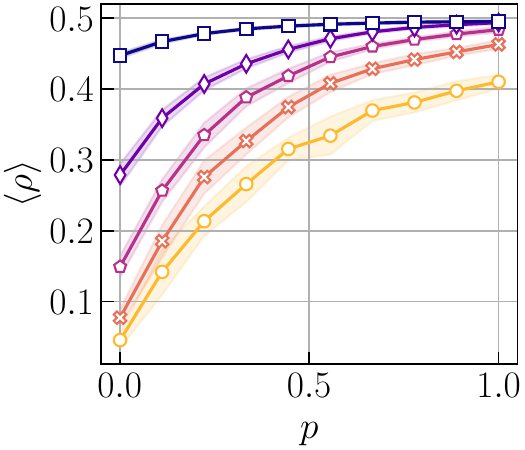}
		\caption{With homophily}
	\end{subfigure}~
	\begin{subfigure}[t]{.23\textwidth}
		\centering
		\includegraphics[width=\textwidth]{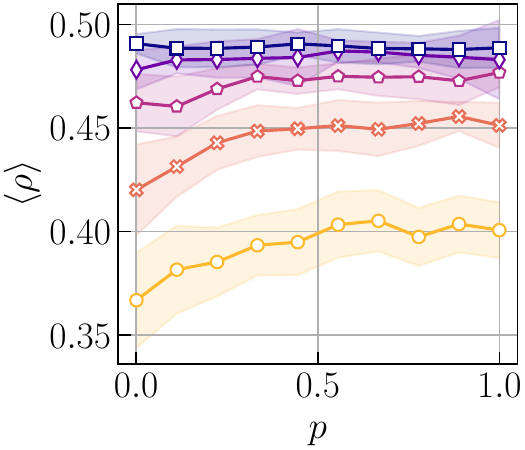}
		\caption{Without homophily}
	\end{subfigure}~
	\begin{subfigure}[t]{.23\textwidth}
		\centering
		\includegraphics[width=\textwidth]{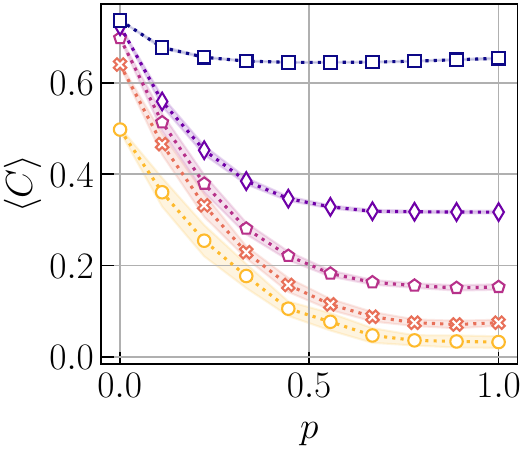}
		\caption{Clustering coefficient}
	\end{subfigure}~
	\begin{subfigure}[t]{.23\textwidth}
		\centering
		\includegraphics[width=\textwidth]{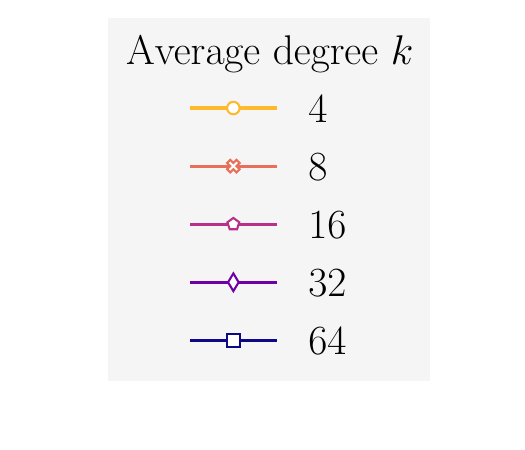}
	\end{subfigure}
	\caption{Impact of clustering coefficient on $\langle\rho\rangle$. For each set of parameters $(k,p)$ we generate $m=30$ Watts-Strogatz networks of size $N=100$, average degree $k$ and rewiring probability $p$. We average $\langle\rho\rangle$ over all realisations. Shaded areas cover $\pm 1$ standard deviation. Half of the agents support opinion 0 $(z^{(0)}>0)$, the other half opinion 1 $(z^{(1)}>0)$. \textbf{(a)} With homophily: the majority of edges are between agents supporting the same opinion. \textbf{(b)} Without homophily: no correlation between edges and opinion support. \textbf{(c)} Average clustering coefficient (independent from homophily). Note the different scales on the y-axes.}
	\label{WSplot}
\end{figure*}

We assume that half of the agents ``support'' opinion 0: they have $z^{(0)}>0$ with a value chosen uniformly at random uniform in $[0,1]$, and $z^{(1)}=0$. The other half analogously supports opinion 1. We consider two cases.
\begin{description}
	\item[With homophily] agents 1 to 50 support opinion 0, and agents 51 to 100 support opinion 1. Thus, most edges connect similar-minded agents.
	\item[Without homophily] supporters of each opinion are placed at random in the graph. Hence, we expect about 50\% of edges to connect similar-minded agents, and 50\% to connect opposite-minded ones.
\end{description}
Results are presented in Fig.~\ref{WSplot}. We also plot the evolution of the average local clustering coefficient $\langle C\rangle$. As expected, $\langle C\rangle$ decreases with $p$ for all values of the average degree $k$. We also observe that $\langle\rho\rangle$ always increases with $k$. This might be due to the fact that, as degrees increase, each neighbor of a node $i$ is less often copied by $i$, meaning less opportunities for them to agree.

With homophily, discord increases significantly with $p$. Agents are initially mostly connected with peers who share the same beliefs, but as we rewire more and more edges, two things happen. First, connections with opposite-minded agents increase as those with similar-minded ones decrease. Second, the clustering coefficient, and thus the number of triangles, fall. These two effects consequently increase exposure to a higher diversity of opinions, which entails higher discord overall.

Without homophily, $\langle\rho\rangle$ also increases with $p$, but much less so. Thus, when there is no homophily in connections, the amount of triangles in the network has a smaller impact on discord. As the community memberships are initially random, rewiring edges has not much impact on the diversity of opinions that agents are exposed to. We also observe a higher variance in $\gald$, which might be due to the higher entropy in the topological distribution of $(z^{(0)},z^{(1)})$. Finally note that, as $p$ approaches $1$, the difference between \textbf{(a)} and \textbf{(b)} vanishes.

\subsection{Discord and communities}

\begin{figure*}
	\centering
         \includegraphics[width=\textwidth]{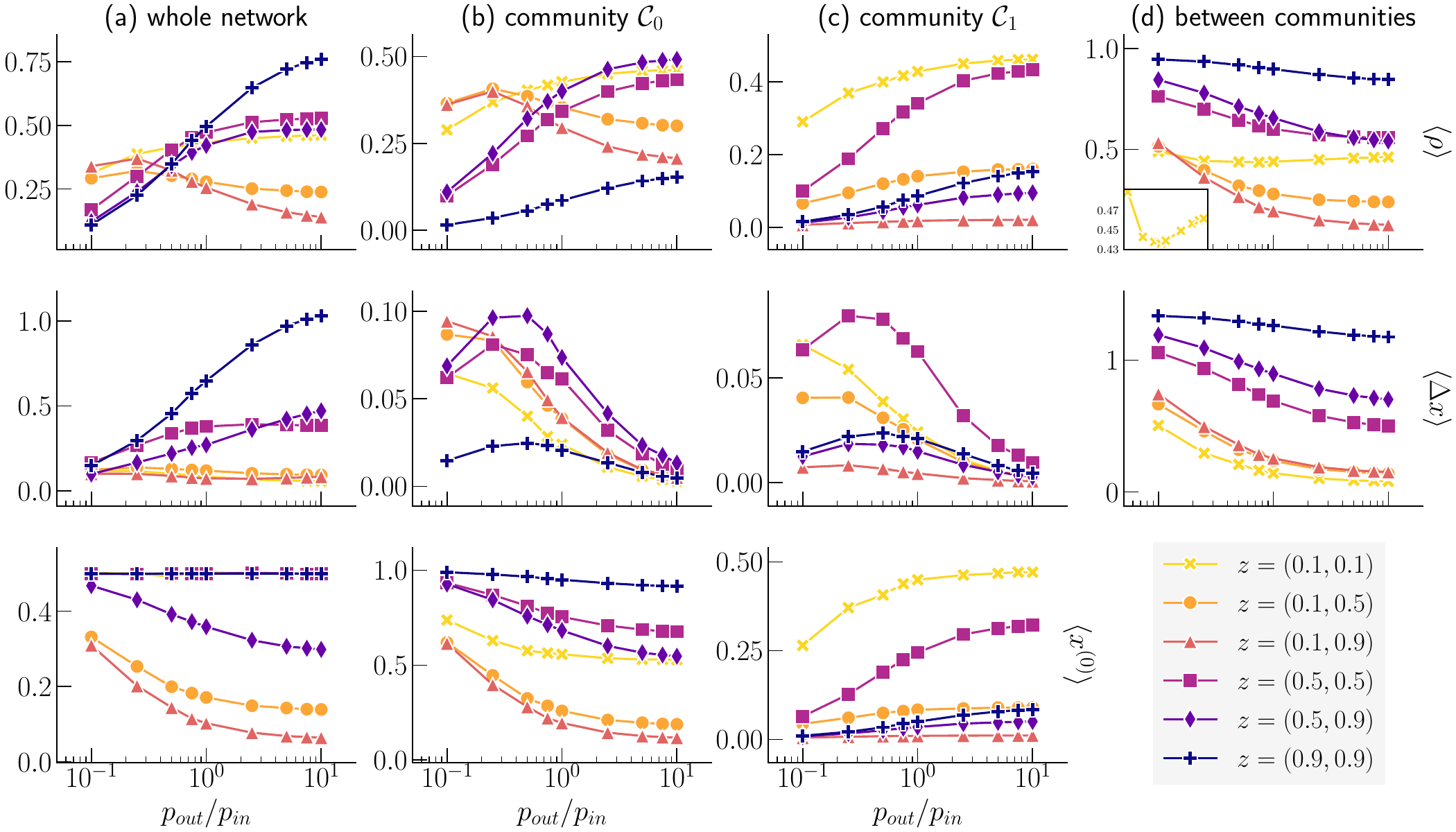}
     \caption{Network of $N=100$ agents with two communities $\mathcal{C}_0$ and $\mathcal{C}_1$. The $s$-zealot exert a total influence $z^s$ on each agent in $\mathcal{C}_s$ and none on others. In-group edge probability is fixed at $p_{in}=0.1$, while the out-group edge probability $p_{out}$ and zealousness $z=(z^0,z^1)$ vary. Results are averaged over 20 agent graphs generated under the Stochastic Block Model, for the whole network (a), within each community (b,c) and between them (d). \textbf{Top:} generalized active links density. \textbf{Middle:} Opinion difference. \textbf{Bottom:} Support for opinion 0. The plots have different scales for clarity, but the purpose is to focus on the qualitative dynamics rather than exact values.}
     \label{plot_sbm}
\end{figure*}

An immediate application of discord lies in the analysis of polarised networks. If two groups support different ideas, how fiercely do they disagree? To what extent does it depend on the connections between them, and on the influence of zealots in each camp? To study these questions we generate networks partitioned into two communities $\mathcal{C}_0$ and $\mathcal{C}_1$ via the Stochastic Block Model, and calculate $\gald$ at equilibrium for various values of the model parameters. The rest of this section is dedicated to the discussion of the results, illustrated in Fig.~\ref{plot_sbm}.

When reinforcing connections between different-minded communities, the novel paths of influence create more diverse information flows, and thus a higher exposition to contradicting beliefs. There are two main consequences one could expect from this. Agents may revise their viewpoint to incorporate adverse ideas, leading the system towards a more consensual state of lower discord. Alternatively, they may fiercely cling onto their prexisting opinions, thus creating more tension and reinforcing discord---the so-called \emph{backfire effect} \cite{bail2018}.

When community $\mathcal{C}_0$ is much less zealous than $\mathcal{C}_1$, for $z=(0.1,0.5)$ and $z=(0.1,0.9)$, outgroup edges can surprisingly reduce the discord within $\mathcal{C}_0$. While adding too few of them will introduce more discord in the community, once they reach a critical mass we observe a decrease in $\gald$. This stems from the fact that $\mathcal{C}_1$ is more zealous and thus has fiercer partisans, meaning opinion 1 gets more and more prevalent in the network as connectivity between the two communities goes up---see the support for opinion 0 dropping in the bottom plots. Thus even within $\mathcal{C}_0$, holding opinion 1 guarantees more agreement with peers. 

It is striking that contrary to $\gald$, the average difference in opinion $\langle \Delta x\rangle$ (Fig.~\ref{plot_sbm}b,c) always decreases within communities, given enough outgroup edges. The opinion difference between $i$ and $j$ is defined by the distance $\Delta x_{ij} = \Vert x_i-x_j\Vert$. Hence, despite a surge of discord amongst like-minded agents, the distributions of their equilibrium opinions converge. This might be due to the fact that as edges are added, the network gets closer to a complete one. Individual node particularities thus fade as agents become more similar to one another. But as the distributions of opinions approach the uniform distribution, the probabilities of drawing two different values increase. This highlights the importance of distinguishing between dissimilarity of opinion distributions and discord probabilities.

Between the communities, we observe that discord always diminishes with more outgroup edges. However, as shown in the inset (Fig.~\ref{plot_sbm}d), for equally low values of zealousness $z=(0.1,0.1)$, first $\gald$ decreases but it quickly goes back up as more outgroup edges are added. Thus, when agents are not too zealous, relations with opposite minded-others become more confrontational as connections between them increase. For high values of $z$, discord does decrease with the number of outgroup edges, but stays important compared with low zealousness cases.

Finally, Fig.~\ref{plot_sbm}a showcases the importance of using other measures than simply average opinions, as also stressed by \cite{vazquez2008}. When the zealousness is the same on both sides, the average opinion $\langle x^{(0)} \rangle$ over the whole network does not change with more edges, but we observe various behaviors in the evolution of $\gald$ and $\langle\Delta x\rangle$.

\section{Conclusion} \label{conclusion}
We developed a novel method to compute discord probabilities in the multi-state voter model on complex networks, with and without zealots. Our results do not rely on any approximation, they are exact and applicable to any directed, weighted network of agents with any finite number of opinions, and individual update rates. This led us to propose a generalized definition of the order parameter $\langle \rho \rangle$ traditionally known as active links density, that quantifies the average discord over all edges. Our new definition accounts for long-range, weighted interactions between agents.

We illustrated our findings on real-life and synthetic networks. We uncovered correlations between discord, edge weights and similarity of ancestors. Through numerical experiments on Watts-Strogatz networks, we observed that clustering had significant impact on discord in the presence of homophily, but less so when that feature was absent. We also found that increasing the number of connections between antagonistic communities could potentially increase discord, depending on the zealousness of agents and density of edges. To gain a deeper understanding of how discord evolves in diverse scenarios, future research shall experiment with various network topologies. Additionally, the theoretical advancements presented here could assist empirical studies examining opinion dynamics in real-world datasets.

\begin{acknowledgments}
	The authors have no competing interests to declare. This project was funded by the UK EPSRC grant EP/S022503/1 that supports the Centre for Doctoral Training in Cybersecurity delivered by UCL's Departments of Computer Science, Security and Crime Science, and Science, Technology, Engineering and Public Policy.
	
	The authors would like to address a special thanks to Effrosyni Papanastasiou for her precious help during the writing process.
\end{acknowledgments}

\section*{Appendices}

\begin{appendix}

\section{Unicity of a solution to Eq.\ 16} \label{unicity} 
We demonstrate that the spectral radius of $V$ is strictly less than 1, which implies that Eq.\ 16 has a unique solution. Let us first introduce a technical lemma from \cite{azimzadeh2018}.
\begin{lemma}[Azimzadeh, 2018, Lemma 2.1 \cite{azimzadeh2018}] \label{spectral_lemma}
	Let $A$ be the adjacency matrix of a graph $\mathcal{F}$ so that $a_{ij}$ is the weight of the edge $j\rightarrow i$. The spectral radius of $A$ is strictly less than 1 if and only if for every row $i$, one of the following holds:
	\begin{itemize}
		\item row $i$ sums to strictly less than 1, or
		\item there is a path $k\rightarrow\ldots\rightarrow i$ in $\mathcal{F}$ and row $k$ sums to strictly less than 1.
	\end{itemize}
\end{lemma}
The matrix $V$ can be seen as the adjacency matrix of a new graph $\mathcal{F}$. Its nodes correspond to agent pairs, and there is an edge from node $i'j'$ to node $ij$ if and only if one of $i'$ or $j'$ is a leader of $i$ or $j$ in the original agent graph $\mathcal{G}$. Let $\nu_{ij}$ denote the sum of the row of $V$ that corresponds to node $ij$. We have
\begin{equation}
	\nu_{ij} = \frac{1}{2} \left( \sum_{k\in\mathcal{L}_i} w_{ik} + \sum_{k\in\mathcal{L}_j}  w_{jk} \right).
\end{equation}
Lemma~\ref{spectral_lemma} tells us that it suffices to prove, for every $ij$: either $\nu_{ij}<1$, or there exists another node $i'j'$ with $\nu_{i'j'}<1$ and a path from $i'j'$ to $ij$ in $\mathcal{F}$. If $\nu_{ij}=1$, assuming every agent can be influenced by a zealot, there exists an agent $k$ such that $z_k^s>0$ for some $s$ and a path from $k$ to $i$. Hence there is path from $ik$ to $ij$ in $\mathcal{F}$, and $\nu_{ik}<1$ as shown by Eq.\ 2 in the main text.

\section{Conditions for independence} \label{independence} 
We now prove that the opinions of $i$ and $j$ are independent if one of the following holds:
\begin{enumerate}
	\item $\sigma_i$ or $\sigma_j$ is constant, or
	\item there is no path from $i$ to $j$ nor from $j$ to $i$, and $i$ and $j$ have no common ancestor.
\end{enumerate}
The first comes from the fact that a constant is independent from any other random variable. It applies in the case one of $i$ and $j$ can be reached by only one zealot, and thus never change opinion. For the second, recall that the opinion distribution of agent $i$ evolves according to
\begin{equation}
	\frac{dx_i^s}{dt} = \sum_{k\in\mathcal{L}_i} w_{ik}x_k^s +z_i^s -x_i^s, \quad s=1,\ldots,S.
\end{equation}
Thus, this evolution is determined by the connections between $i$ and zealots, and the opinion distributions of the leaders of $i$. In turn, the opinion distribution of a leader $k$ of $i$ is function of the connections between $k$ and zealots, and the opinion distributions of the leaders of $k$. Iterating this reasoning, eventually all ancestors of $i$---and only them---intervene. Thus, if $j$ is an ancestor of $i$, the opinions of the two are not independent. Otherwise, if there is no path between them but they have a common ancestor, their opinions are both function of the opinion of that ancestor, meaning they are not independent. If neither $i$ nor $j$ is an ancestor of the other, and if they have no common ancestor, their opinions are two independent random processes.

\section{The datasets} \label{datasets} 
\begin{figure}
	\centering
         \includegraphics[width=.45\textwidth]{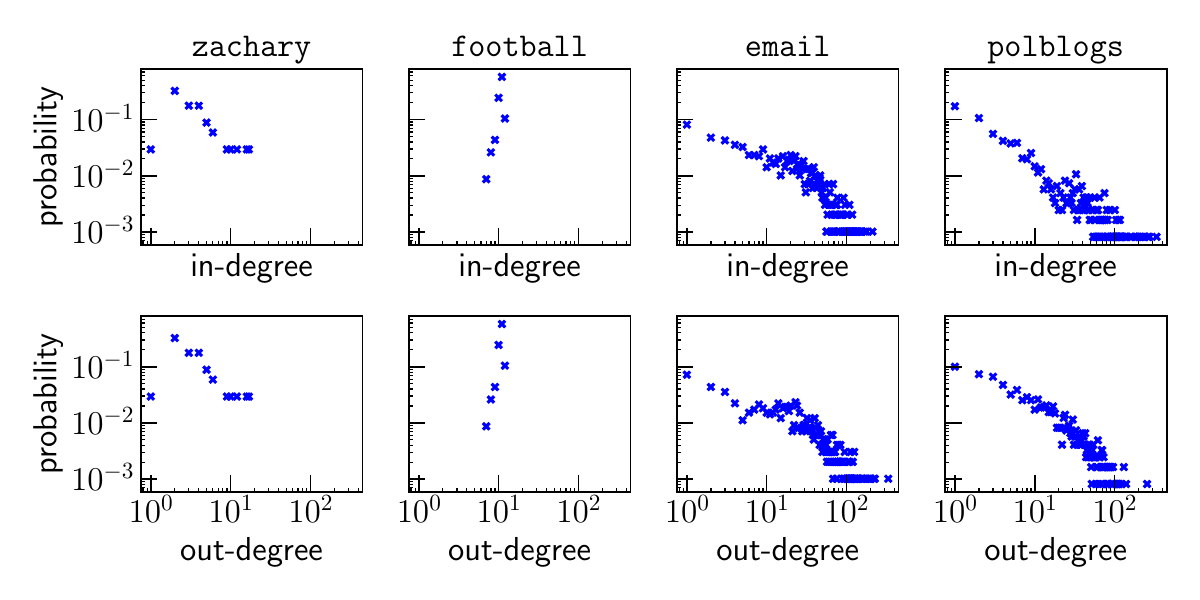}
     \caption{Unweighted degree distributions for each toy dataset, logarithmic scale. \texttt{Zachary} and \texttt{Football} are undirected so both distributions are the same. \textbf{Top:} in-degrees. \textbf{Bottom:} out-degrees.}
     \label{deg_distrib}
\end{figure}

In Sect.~\ref{dependency} we use four datasets, described in Table~\ref{data_stats}. In each of them, a node belongs to a single community, given by the creators of the dataset. We only keep the largest weakly connected component of each network. The in-degree and out-degree distributions we obtain are presented in Fig.~\ref{deg_distrib}. 

\begin{table*}[htbp!]
	\centering
	\renewcommand{\arraystretch}{.65}
	\begin{tabular}{lcccc}
	& Zachary  \cite{zachary} & Football \cite{football} & Email \cite{email} & Polblogs \cite{polblogs} \\ \hline
	\textbf{Number of agents} & 34 & 115 & 986 & \textbf{1,222} \\
	\textbf{Number of edges} & 156 & 1,226 & \textbf{25,552} & 19,021 \\
	\textbf{Density} & \textbf{0.14} & 0.09 & 0.03 & 0.01 \\
	\textbf{Directed} & No & No & Yes & Yes \\
	\textbf{Self-loops} & No & No & Yes & No \\
	\textbf{Mean degree} & 4.59 & 10.66 & \textbf{25.91} & 15.57 \\
	\textbf{Number of communities} & 2 & 12 & \textbf{42} & 2 \\
	\textbf{Modularity} & 0.36 & \textbf{0.53} & 0.31 & 0.41 \\
	\textbf{Average clustering} & \textbf{0.57} & 0.40 & 0.41 & 0.32 \\
	\textbf{Independent agent pairs} & 0\% & 0\% & 4\% & \textbf{40\%} \\ \hline
	\end{tabular}
		\caption{Basic statistics for the datasets. Modularity and average clustering are computed on the undirected, unweighted networks. See Sect.~\ref{indep_pairs} for a definition of independent agent pairs.}
		\label{data_stats}
\end{table*}

The datasets are initially unweighted. We first set $w_{ij}$ to 1 if there is an edge from $j$ to $i$, and zero otherwise. Once all such values are set, we define values of zealousness as follows: for any agent $i$ who belongs to community $s$, we attribute a random uniform value between 0 and 1 to $z_i^s$, and set $z_i^r=0$ for $r\neq s$. We then proceed to multiply each weight $w_{ij}$ by
\begin{equation}
	\frac{1-z_i^s}{\sum_{k\in\mathcal{L}_i} w_{ik}},
\end{equation}
with $s$ the community of user $i$, so that Eq.~\ref{normalisation} holds. To highlight the absence of approximation in our results, we study the difference between empirical values of discord obtained in simulation, and theoretical ones from Eqs.~\ref{qij_indep} and \ref{qij_dep}. For the simulation we proceed as per Algorithm~\ref{simu_algo}, with $T=10^5N$ steps and a burn time of $T_b=10N$ steps. This means the first $10N$ steps are not taken into account when computing discord probabilities, in order to reduce the influence of the initial state. 

\begin{figure}
	\centering
	\begin{subfigure}[b]{.225\textwidth}
		\centering
		\includegraphics[width=\textwidth]{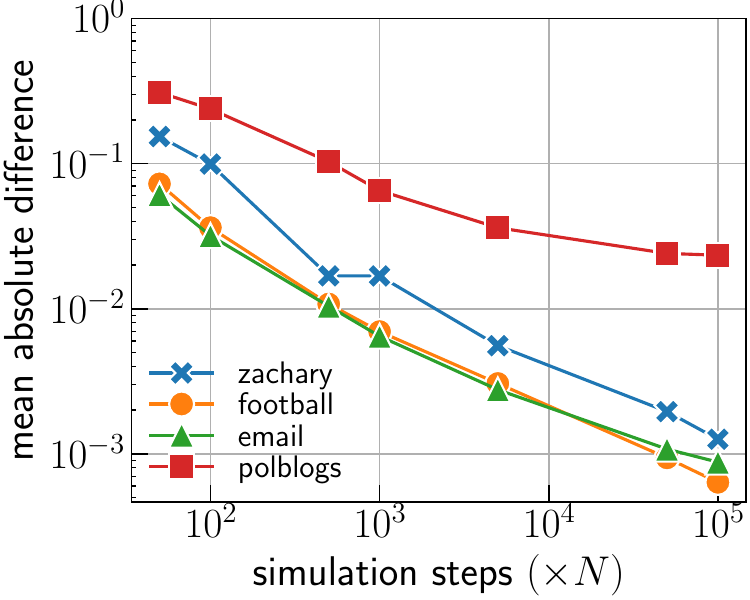}
	\end{subfigure}~
	\begin{subfigure}[b]{.225\textwidth}
		\centering
		\includegraphics[width=\textwidth]{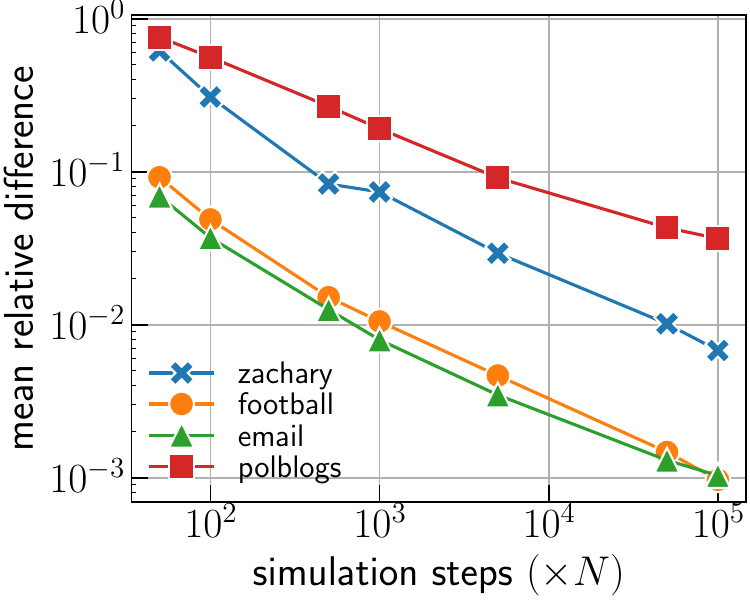}
	\end{subfigure}
	\caption{Difference between simulated and theoretical values of discord.}
	\label{simuVStheo}
\end{figure}

Simulated values are compared with their theoretical counterparts in \autoref{simuVStheo}. For each pair $(i,j)$ we compute the absolute and relative differences, respectively $\vert\rho_{ij}^\text{theo}-\rho_{ij}^\text{simu}\vert$ and $\vert\rho_{ij}^\text{theo}-\rho_{ij}^\text{simu}\vert/\rho_{ij}^\text{theo}$. We then average these values over all agent pairs. As the simulation time increases, we observe a tighter and tighter fit between the values, with relative differences reaching between $10^{-3}$ and $10^{-2}$ for all four datasets---except for \texttt{polblogs}, with a difference slightly below $10^{-1}$. This highlights the exactness of the equations we derived. 

\begin{algorithm}
	\DontPrintSemicolon
	\KwData{user set $\mathcal{N}$, number of opinions $S$, adjacency matrix $W$, zealousness matrix $Z$, number of steps $T$, burn time $T_b$}
	\KwResult{estimated discord probabilities $\rho_{ij}$ for all user pairs $(i,j)$}
	\Begin{
	$x_i \sim \mathcal{U}_{\{1,\ldots,S\}}$ for all $i\in\mathcal{N}$ \tcp*[f]{initialize users opinions at random}\;
	$\rho_{ij} \longleftarrow 0$ for all $i,j\in\mathcal{N}$ \tcp*[f]{initialize probabilities to zero}\;
	$\tau_{ij}\longleftarrow T_b+1$ for all $i,j\in\mathcal{N}$ \tcp*[f]{time of last update for each pair}\;
	\For{$1\le t\le T$}{
		$i \sim \mathcal{U}_{\{1,\ldots,N\}}$ \tcp*[f]{select a random user}\;
		$x_i^\text{old} \longleftarrow x_i$  \tcp*[f]{store current opinion}\;
		$x_i \longleftarrow x_j$ with probability \ $w_{ij}$, or $s$ with probability\ $z_i^s$  \tcp*[f]{draw new opinion}\;
		\If{$t>T_b$ and $x_i\neq x_i^\text{old}$}{
			\tcp{if $i$ changed opinion, for users $j$ who were disagreeing with $i$ until now, we add the duration of the disagreement to $\rho_{ij}$}
			\For{$j\in\mathcal{N}\backslash\{i\}$}{
				\If{$x_j\neq x_i^\text{old}$}{
					$\rho_{ij} \longleftarrow \rho_{ij}+(t-\tau_{ij})$ \tcp*[f]{update discord probability}\;
					$\tau_{ij} \longleftarrow t$ \tcp*[f]{set time of last update to current time}\;
				}
			}
		}
	}
	$\rho_{ij} \longleftarrow \rho_{ij}/(T-T_b)$ for all $i,j\in\mathcal{N}$ \tcp*[f]{normalize probabilities}\;
	}
	\caption{Estimation of discord probabilities via simulation}
	\label{simu_algo}
\end{algorithm}

\end{appendix}

	\bibliographystyle{apsrev4-2}
	\bibliography{biblio}

\end{document}